\documentclass[showpacs,prb,amsmath,amssymb,superscriptaddress,nobibnotes,twocolumn, preprintnumbers]{revtex4-1}
\usepackage{graphicx}
\usepackage{endnotes}
\usepackage{bm}
\usepackage{epsfig}
\usepackage{amsmath}

\newcommand{\ie}{{\it i.e.}}

\newcommand{\etal}{{\it et al.}}
\newcommand{\SrIr}{Sr$_{3}$Ir$_4$Sn$_{13}$}
\newcommand{\SrRh}{Sr$_{3}$Rh$_4$Sn$_{13}$}

\newcommand{\CaSrRhx}{(Ca$_{x}$Sr$_{1-x}$)$_3$Rh$_4$Sn$_{13}$}
\newcommand{\CaSrIrx}{(Ca$_{x}$Sr$_{1-x}$)$_3$Ir$_4$Sn$_{13}$}

\newcommand{\CaIr}{Ca$_{3}$Ir$_4$Sn$_{13}$}
\newcommand{\LaSn}{La$_{3}$Co$_4$Sn$_{13}$}
\newcommand{\Tstar}{$T^*$}
\newcommand{\Tc}{$T_c$}

\begin{document}

\title{Second-order Structural Transition in Superconductor La$_3$Co$_4$Sn$_{13}$}
\author{Y. W. Cheung}
\author{J. Z. Zhang}
\author{J. Y. Zhu}
\author{W. C. Yu}
\author{Y. J. Hu}
\author{D. G. Wang}
\affiliation{Department of Physics, The Chinese University of Hong Kong, Shatin, New Territories, Hong Kong, China}

\author{Yuka Otomo}
\author{Kazuaki Iwasa}
\altaffiliation{Present address: Frontier Research Center for Applied Atomic Sciences, Ibaraki University, Shirakata 162-1, Tokai, Naka, Ibaraki 319-1106, Japan}
\affiliation{Department of Physics, Tohoku University, Sendai 980-8578, Japan}

\author{Koji Kaneko}
\affiliation{Materials Sciences Research Center, Japan Atomic Energy Agency, Tokai, Naka, Ibaraki 319-1195, Japan}

\author{Masaki Imai}
\author{Hibiki Kanagawa}
\affiliation{Department of Chemistry, Graduate School of Science, Kyoto University, Kyoto 606-8502, Japan}
\author{Kazuyoshi Yoshimura}
\affiliation{Department of Chemistry, Graduate School of Science, Kyoto University, Kyoto 606-8502, Japan}
\affiliation{Research Center for Low Temperature and Materials Sciences, Kyoto University, Kyoto 606-8501, Japan}

\author{Swee K. Goh}
\email{skgoh@phy.cuhk.edu.hk}
\affiliation{Department of Physics, The Chinese University of Hong Kong, Shatin, New Territories, Hong Kong, China}
\date{\today}

\begin{abstract}
The quasi-skutterudite superconductor La$_3$Co$_4$Sn$_{13}$ undergoes a phase transition at $T^*=152$~K. By measuring the temperature dependence of heat capacity, electrical resistivity, and the superlattice reflection intensity using X-ray, we explore the character of the phase transition at $T^*$. Our lattice dynamic calculations found imaginary phonon frequencies around the ${\bf M}$ point, when the high temperature structure is used in the calculations, indicating that the structure is unstable at the zero temperature limit. The combined experimental and computational results establish that $T^*$ is associated with a second-order structural transition with $\bm{q}$=(0.5,~0.5,~0) (or the {\bf M} point). Further electronic bandstructure calculations reveal Fermi surface sheets with low curvature segments, which allow us to draw qualitative comparison with both Sr$_3$Ir$_4$Sn$_{13}$ and Sr$_3$Rh$_4$Sn$_{13}$ in which similar physics has been discussed recently.

\end{abstract}

\pacs{74.25.fc, 74.25.Bt, 63.20.kd}


\maketitle

\section{Introduction}
Superconducting stannides \cite{Remeika1980,Espinosa1980} with stoichiometry $A_3T_4$Sn$_{13}$ ($A$=La, Sr, Ca, $T$=Co, Rh, Ir) have received a renewed attention \cite{Klintberg2012,Goh2015,Yu2015,Kuo2014,Kuo2015,Yang2010, Gerber2013,Liu2013,Slebarski2014,Fang2014,BChen2015,Mazzone2015,Wang2012,Biswas2014,Wang2015, Kase2011, Hayamizu2011, Zhou2012, Sarkar2015,Biswas2015, Slebarski2013, Thomas2006, Slebarski2015,Neha2016,Tompsett2014, XChen2015} owing to the discovery of a structural transition which can be tuned to 0~K \cite{Klintberg2012,Goh2015,Yu2015}. In \SrIr\ and \SrRh, the structural transition occurs at \Tstar$\simeq147~$K \cite{Klintberg2012,Kuo2014} and 138~K \cite{Goh2015,Kuo2015}, respectively. In these systems, pronounced anomaly can be seen at \Tstar\ in various physical properties \cite{Yang2010,Gerber2013,Liu2013,Slebarski2014,Fang2014,Kuo2014,Klintberg2012,Goh2015,Kuo2015,Yu2015,BChen2015,Wang2012,Biswas2014,Wang2015,Mazzone2015}, including electrical resistivity, magnetic susceptibility and specific heat. With applied pressure or the substitution of Sr by Ca, \ie\ \CaSrIrx\ and \CaSrRhx, \Tstar\  decreases rapidly, accompanied by a moderate increase in the superconducting transition temperature, \Tc, which peaks near the composition/pressure where \Tstar\ extrapolates to 0~K. The phase diagrams constructed thus highly resemble the ones constructed for many topical superconductors found in the vicinity of a magnetic quantum critical point \cite{Mathur1998,Gegenwart2008,Paglione2010,Ishida2009,Hashimoto2012,Shibauchi2014}.

In the \SrIr\ and \SrRh\ systems, the superconducting gap function is nodeless \cite{Kase2011,Hayamizu2011,Zhou2012,Wang2012,Biswas2014,Wang2015,Sarkar2015} and hence the superconductivity is of the conventional $s$-wave type. In the vicinity of the putative structural quantum critical point where \Tstar\ extrapolates to 0~K, $\mu$SR \cite{Biswas2015} and specific heat \cite{Yu2015} detected a strongly enhanced electron-phonon coupling strength, indicated by the enhancement in $2\Delta(0)/k_BT_c$ and $\Delta C/\gamma T_c$ beyond the BCS weak-coupling values\cite{BCS1,BCS2,Poole2007}, where $\Delta(0)$ is the size of the gap, $\Delta C$ is the specific heat jump at \Tc\ and $\gamma$ is the Sommerfeld coefficient. Therefore, the \SrIr\ and \SrRh\ systems, as well as the corresponding substitution series, are promising materials for exploring the interplay between structural instability and strong-coupling superconductivity.

\LaSn\ is one of the compounds in the family isostructural to \SrIr\ and \SrRh\ at room temperature (space group $Pm\bar{3}n$) \cite{Slebarski2013,Thomas2006}. It is a superconductor with a \Tc\ of $\sim2.7$~K \cite{Slebarski2015,Thomas2006}. Incidentally, \LaSn\ has been shown to exhibit a feature at $T^*\sim152$~K, detectable in electrical resistivity, heat capacity, NMR Knight shift, and nuclear spin-lattice relaxation rate \cite{Slebarski2014, Liu2013}. Based on X-ray diffraction \cite{Slebarski2013} and the hysteresis in the specific heat \cite{Liu2013}, Liu \etal\ argued that the feature at \Tstar\ is associated with a {\it first-order} structural phase transition. However, Neha \etal\ did not observe the feature in their sample \cite{Neha2016}. Furthermore, using first principle calculations, they showed that all phonon mode frequencies are real, indicating that the room temperature structure with the space group $Pm\bar{3}n$ is stable and hence there should not be a structural transition.

High pressure transport studies show that \Tstar\ in \LaSn\ decreases to $\sim115$~K at 25~kbar \cite{Slebarski2014}. If this trend continues to higher pressures, it can give rise to a  structural quantum phase transition similar to the cases of \SrIr\ and \SrRh\ under pressure. If the phase transition remains second-order at zero temperature, a structural quantum critical point can be realised. Therefore, understanding the nature of the phase transition at \Tstar\ is crucial. In this manuscript, we revisit the problem and study our own single crystals of \LaSn, which is designed to settle the dispute. Our results from electrical resistivity, specific heat, X-ray diffraction as well as density functional theory calculations support the scenario that the phase transition at \Tstar\ is a {\it second-order} structural phase transition.

\section{Method}

Single crystals of La$_3$Co$_4$Sn$_{13}$ were obtained by a tin flux method.
Elemental La(3N), Co(3N), and Sn(3N) were sealed in quartz tubes in the ratio of 2: 1: 30, and then heated up to 1050~$^\circ$C and slowly cooled to 600~$^\circ$C at the rate of 3$^\circ$C/hr. Excess tin was removed by centrifugation after reheating to 500~$^\circ$C, and then removed by dilute hydrochloric acid. Excellent homogeneity was confirmed using Oxford Instruments X-MAX 50 energy dispersive X-ray detector in JEOL JSM-7800F scanning electron microscope. Heat capacity was measured using a standard pulse relaxation method. The mass of the sample is 24~mg. Electrical resistivity was measured using the four-contact method. A Physical Property Measurement System (Quantum Design) was used to provide the low temperature and high magnetic field environment. X-ray diffraction measurements were performed with a cryostat installed on a conventional four-circle diffractometer equipped with a rotating-anode X-ray generator consisting of a molybdenum target. The $K_\alpha$ X-ray was chosen by using a pyrolytic graphite monochromator. All samples used are from the same batch.

The calculations were based on density functional theory \cite{Hohenberg1964,kohn1965}. The VASP code \cite{kresse1996} with a plane wave basis set \cite{blochl1994,kresse1999} in conjunction with the package of PHONOPY \cite{togo2008,Togo2015} was used to calculate the phonon spectra. Computational details are discussed in Supplemental Material.
The electronic structures were calculated using the all-electron full-potential linearized augmented plane-wave code WIEN2k \cite{schwarz2003}. The muffin-tin radii were set to 2.5 a.u. for the La atoms and 2.37 a.u. for the Co and Sn atoms. $R_{\rm{MT}}^{\rm{min}}K_{\rm{max}}=7$ and a $k$-point mesh of 15000 in the first Brillouin zone were used to achieve convergence in the density of states and electron band structure.

\section{Results and Discussion}
\begin{figure}[!t]\centering
      \resizebox{9cm}{!}{
              \includegraphics{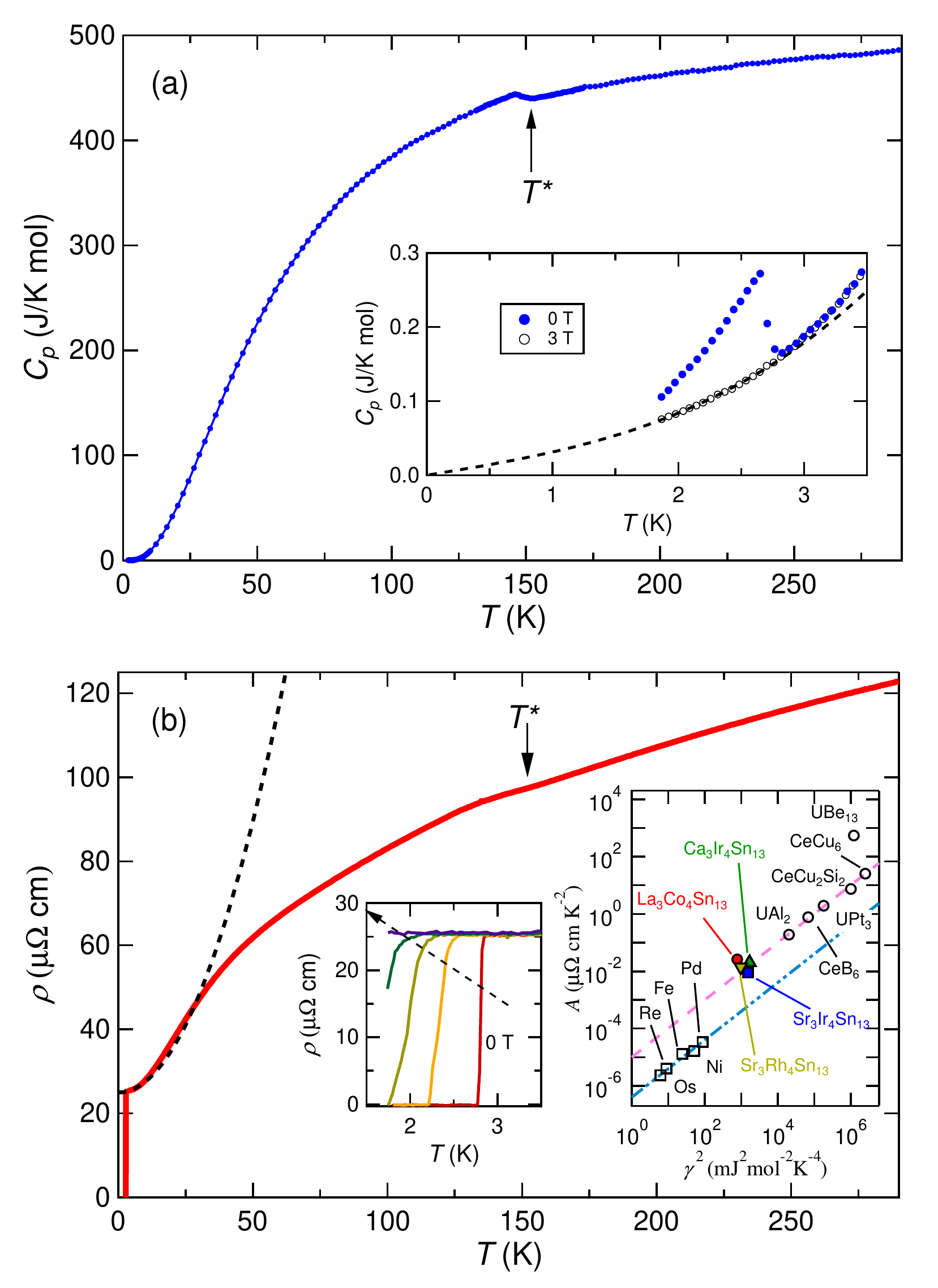}}                				
              \caption{\label{fig1} (Color online) (a) The temperature dependence of the specific heat in \LaSn. The solid arrow indicates \Tstar. Inset: an expanded view of the specific heat around \Tc\ at zero field and 3~T. (b) The temperature dependence of the electrical resistivity. The solid arrow indicates \Tstar and the dotted line shows that the low temperature resistivity follows a Fermi liquid behaviour (see text). Left inset: an expanded view of the low temperature resistivity showing the suppression of $T_c$ with the application of 0, 0.2, 0.4, 0.6, and 1~T. The arrow indicates the direction of an increasing magnetic field. Right inset: $A$-$\gamma^2$ plot for selected heavy fermion compounds (open circles), transition metals (open squares) and several 3-4-13 systems. The dashed and dash-dot lines indicate the Kadowaki-Woods ratio commonly observed in heavy fermion compounds and transition metals, respectively.}
\end{figure}

Fig.~\ref{fig1} displays the overall behaviour of the specific heat, $C_p(T)$, and the electrical resistivity, $\rho(T)$, in \LaSn. A weak feature can be seen at around 152~K, which corresponds to the structural phase transition to be discussed later. We first discuss the low temperature properties of \LaSn. At low temperature, a clear jump in $C_p$ is detected at 2.7~K, which corresponds to the superconducting transition, as evidenced in the rapid disappearance of the electrical resistivity at the same temperature (insets to Fig.~\ref{fig1}). The superconducting phase transition temperature is consistent with previous reports\cite{Thomas2006, Slebarski2015}. When magnetic field is applied, superconductivity can be suppressed. At 3~T, $C_p(T)$ down to the lowest attainable temperature does not exhibit any anomaly associated with superconductivity. By replotting the high field data on the axes of $C_p/T$ against $T^2$ (see Supplemental Material), a linear region up to $T=2.7$~K can be found, with slope $\beta=3.53$~mJ$\cdot$K$^{-4}$mol$^{-1}$ and intercept $\gamma=27.87$~mJ$\cdot$K$^{-2}$mol$^{-1}$. This allows us to extract the electronic contribution $\gamma T$ and the phonon contribution $\beta T^3$, with the sum $\gamma T+\beta T^3$ shown as the dashed line in the inset to Fig.~\ref{fig1}(a). The normalized specific heat jump $\Delta C/\gamma T_c=1.93$, which is larger than the BCS weak coupling value of 1.43.

At low temperature, $\rho(T)$ follows a $T^2$ behaviour, \ie\ $\rho(T)=\rho_0+AT^2$. By plotting $\rho(T)$ vs. $T^2$ (see Supplemental Material), a linear region is found to extend from $T_c$ to 7.2~K, whose slope and intercept give the $A$-coefficient of 0.026~$\mu\Omega$cm$\cdot$K$^{-2}$ and $\rho_0=24.9$~$\mu\Omega$cm, respectively. The dashed line in Fig.~\ref{fig1}(b) is generated with $A$ and $\rho_0$ obtained. From the analysis of the low temperature normal state data, the Kadowaki-Woods ratio A/$\gamma^2$ can be calculated to be 3.3$\times10^{-5}$~$\mu\Omega$cm$\cdot$mol$^2$K$^2$mJ$^{-2}$, which is close to the value commonly observed in  heavy fermion compounds \cite{Kadowaki1986, Jacko2009}. Interestingly, the Kadowaki-Woods ratio of several related 3-4-13 compounds, namely \SrIr \cite{Kuo2014}, \SrRh \cite{Kuo2015}, and \CaIr \cite{Yang2010}, are found to cluster around the same region in the $A$-$\gamma^2$ plot, as shown in the right inset to Fig. 1(b).

\begin{figure}[!t]\centering
       \resizebox{8.5cm}{!}{
              \includegraphics{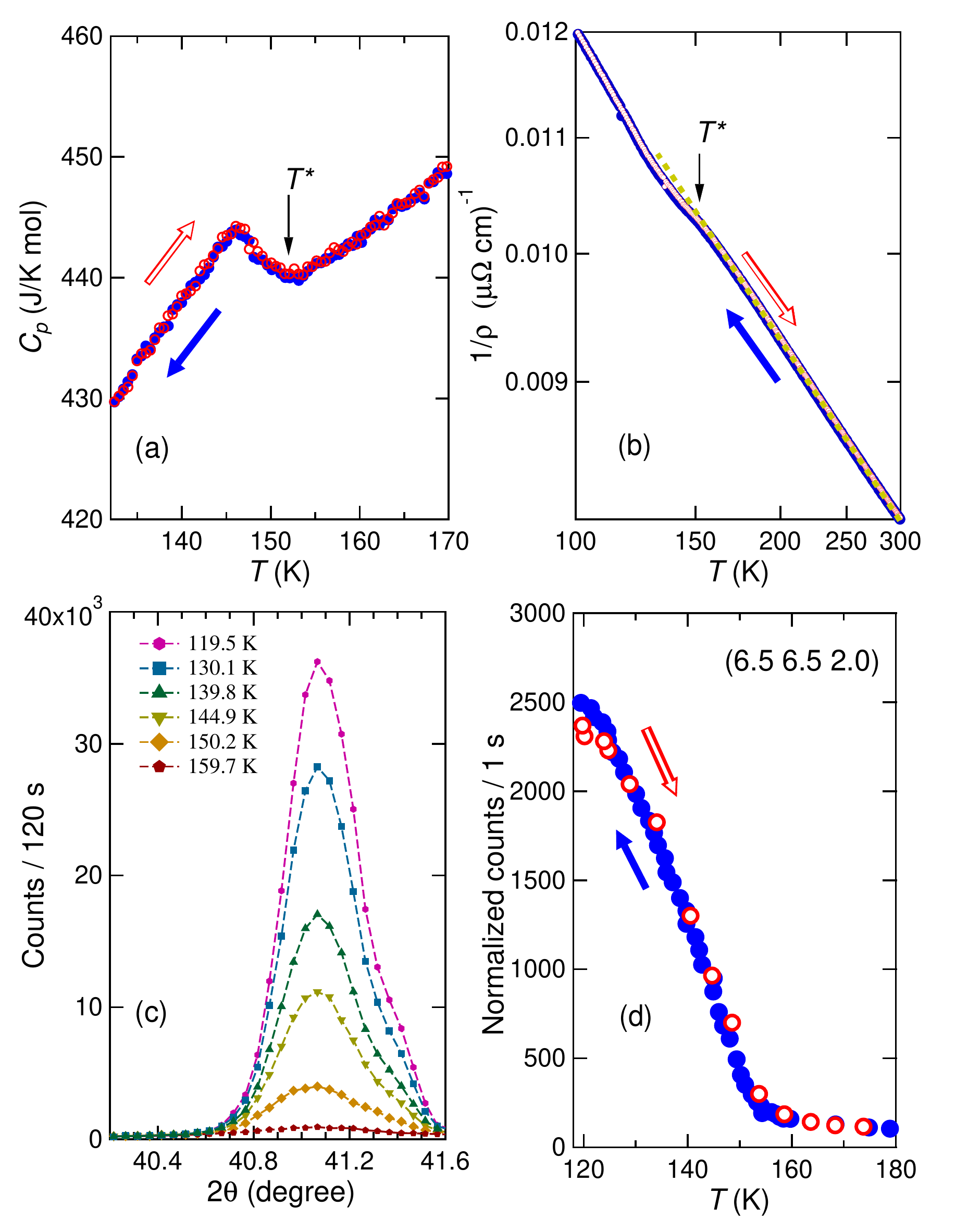}}                				
              \caption{\label{fig2} (Color online) The temperature dependence of (a) specific heat, (b) inverse resistivity around \Tstar. (c) Selected X-ray intensities at ${\bm Q}=(6.5, 6.5, 2.0)$ across \Tstar\ and (d) the temperature dependence of the integrated X-ray intensity at ${\bm Q}=(6.5, 6.5, 2.0)$. Note that in (b), both axes are plotted using logarithmic scale. For all quantities shown in (a), (b), and (d), the data were collected on cooling and on warming, with warm-up (cool-down) data denoted by open (filled) symbols and the open (filled) arrow. 
              }
\end{figure}
We now turn our attention to the high temperature phase transition at \Tstar. Fig.~\ref{fig2}(a) shows the temperature dependence of the specific heat around \Tstar. A clear lambda-like jump in the specific heat is detected. Particular care was taken to measure the specific heat on cooling and on warming. The datasets are displayed in Fig.~\ref{fig2}(a) with the arrows denoting the direction of the temperature sweep. It is clear that no discernable hysteresis exists. This is in stark contrast to the datasets presented by Liu \etal, where they detected a hysteresis as large as $\sim3-5$~K near \Tstar\ in their specific heat data \cite{Liu2013}. Based on the shape of the specific heat jump and the absence of thermal hysteresis, we conclude that the phase transition at \Tstar\ is {\it second-order}.

It is harder to see the effect of phase transition on $\rho(T)$. Following \'{S}lebarski \etal, we replot the resistivity data on the axes $\ln(1/\rho)$ vs. $\ln T$. \Tstar\ is the temperature below which the graph of $\ln(1/\rho)$ vs. $\ln T$ deviates from linearity. Indeed, the value of \Tstar\ extracted from our specific heat data agrees well with the assignment based on the deviation from the linearity, as shown in Fig.~\ref{fig2}(b). For the resistivity, the measurement was again performed on cooling and on warming -- absence of thermal hysteresis is apparent in the data, consistent with the conclusion that the phase transition is second-order we reached from the analysis of the specific heat data.

\begin{figure}[!t]\centering
       \resizebox{8cm}{!}{
              \includegraphics
              {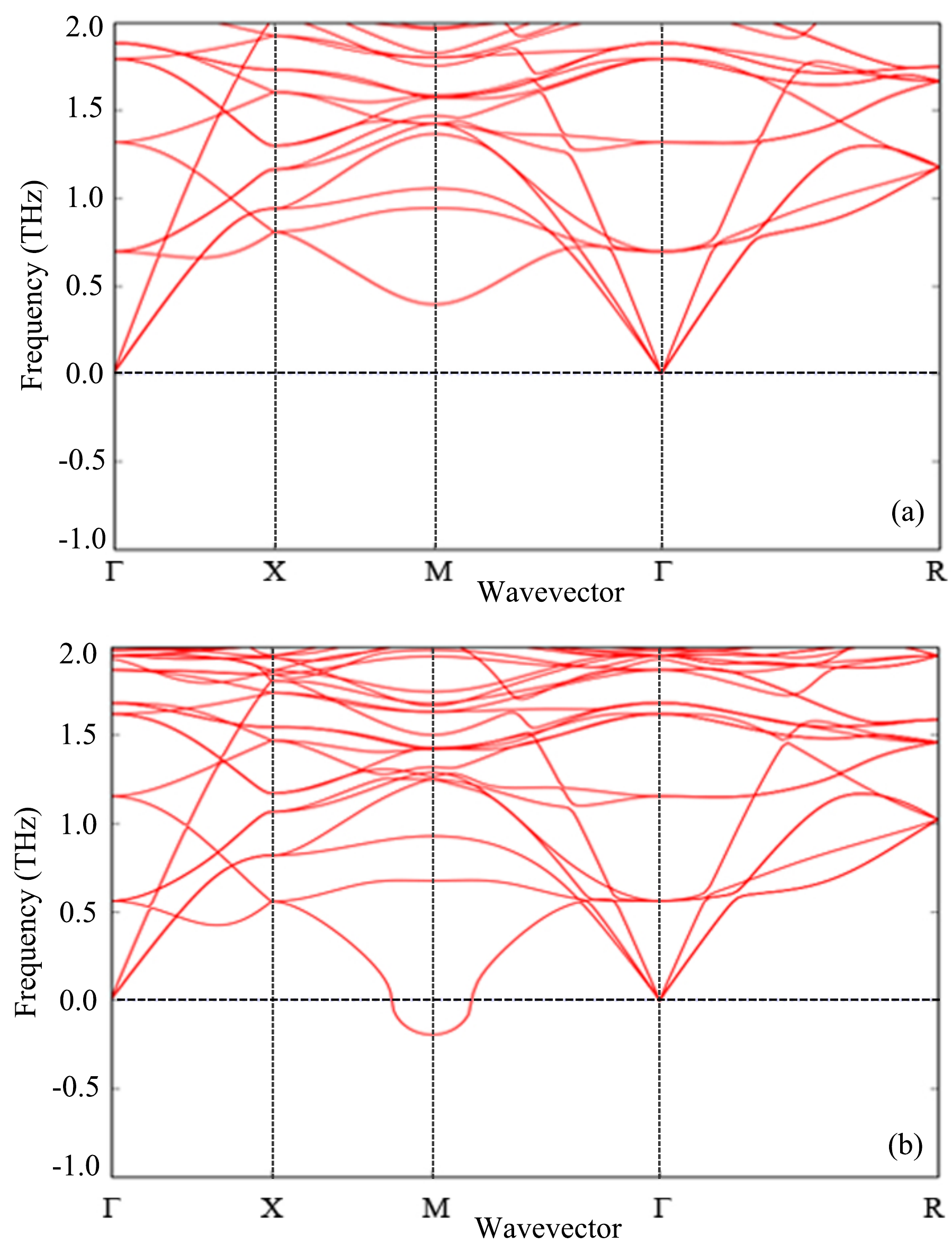}}              				
              \caption{\label{fig3} (Color online) Calculated phonon spectra of \LaSn\ by LDA with (a) the theoretically optimized lattice constant, and (b) the experimental lattice constant. Only the low-lying frequencies are shown.}
\end{figure}

Having established the second-order nature of the \Tstar\ transition, we now proceed to measure the superlattice reflection intensity by using X-ray diffraction. Calculations, to be discussed later, suggest that the structural instability is associated with a modulation vector ${\bm q}=$(0.5, 0.5, 0.0). Therefore, additional reflections with ${\bm Q}={\bm k} + {\bm q}$ should appear below \Tstar, where $\bm{k}$ corresponds to the Bragg spots in the high temperature $Pm\bar{3}n$ phase. Fig.~\ref{fig2}(c) displays the temperature evolution of selected X-ray diffraction intensities for ${\bm Q}=(6.5, 6.5, 2.0)$ across \Tstar. While the intensity of ${\bm k}=(6.0, 6.0, 2.0)$ shows a weak temperature dependence (not shown), the intensity of (6.5, 6.5, 2.0) is negligible above \Tstar\ but it grows continuously below \Tstar, as shown in Fig.~\ref{fig2}(d). Furthermore, the superlattice reflection intensity does not exhibit a discernable thermal hysteresis. All these features thus strengthen the claim that the phase transition at \Tstar\ is continuous.

In order to assess the stability of the high temperature structure with space group $Pm\bar{3}n$ at 0~K, we have performed the lattice dynamic calculations. The optimized lattice constant is 9.4498~{\AA} using LDA as the exchange-correlation functional. A detailed discussion and comparison of \CaIr , \SrIr\ and \LaSn\ are described in Supplemental Material.
Fig.~\ref{fig3}(a) shows the phonon spectrum along the high symmetry directions calculated using the theoretically optimized lattice constant by LDA. A softened branch of phonon modes with a minimum frequency at the $\mathbf{M}$ point can be seen, which corresponds to $\bm{q}=(0.5, 0.5, 0.0)$. This is in contrast to the finding by Neha \emph{et al.} where no softened phonon modes were observed \cite{Neha2016}. Additionally, we also calculated the spectrum using the experimental lattice constant for further comparison \cite{Slebarski2013}. This approach has been adapted in previous works to investigate the lattice instability and were shown to give consistent results \cite{Tompsett2014,Yu1995,Ignatov2003,Ghosez1999,Subedi2013}. From Fig.~\ref{fig3}(b), one can see that the phonon spectrum possesses a branch of imaginary frequencies. The presence of these imaginary frequencies suggests a structural instability which drives the system towards a displacive structural transition and it is consistent with our experimental findings. 




\begin{figure}[!t]\centering
       \resizebox{8.5cm}{!}{
              \includegraphics{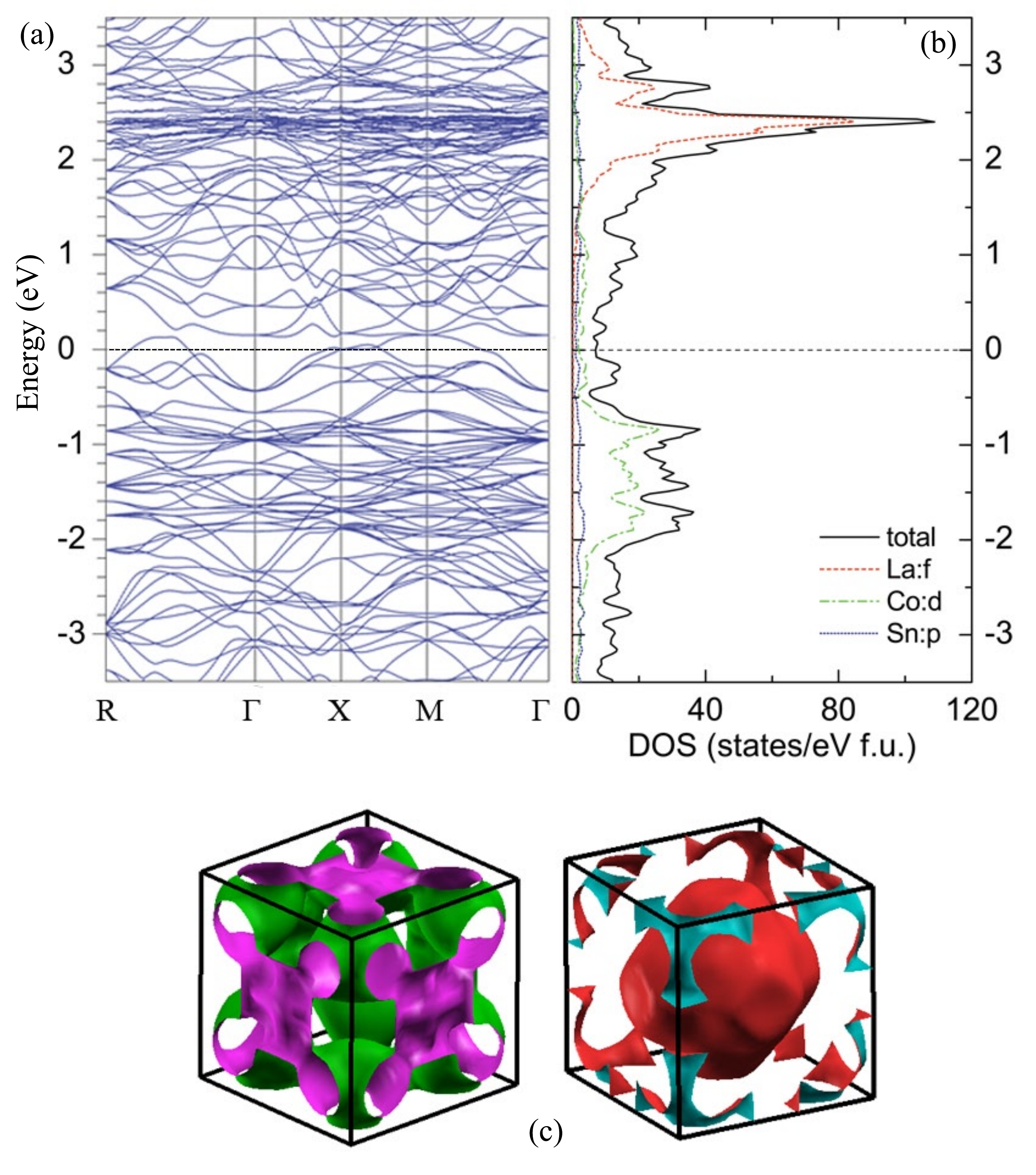}  }           				
              \caption{\label{fig4} (Color online) (a) The electronic band structure of \LaSn\ along the high symmetry directions. (b) The total and partial density of states. The energy scale is shown with respect to the Fermi energy (dashed line). (c) The Fermi surface sheets in the first Brillouin zone. ${\bf \Gamma}$ is at the centre of the cube.}
\end{figure}

Fig.~\ref{fig4} shows the electronic structure of \LaSn\ in the high temperature phase from LDA calculation. There are two dispersive bands crossing the Fermi level indicating the metallic nature of the compound. The La 4f states are empty and is localized around 2.5 eV above the Fermi energy (=0~eV) while the Co 3d states are bounded between $-0.5$~eV and $-2$~eV. The results are similar to that obtained from the GGA+SOC calculation \cite{Zhong2009}. At the Fermi level, the states are mainly contributed by the Co 3d and Sn 5p electrons in the CoSn(2)$_{6}$ trigonal prisms around an Sn(1)Sn(2)$_{12}$ icosahedral cages. The total density of states is $\sim~7.2$~states/eV per formula unit. This translates into a Sommerfeld coefficient $\gamma_{\rm{cal}}$ of about 17.0~mJ$\cdot$K$^{-2}$mol$^{-1}$ under the free electron approximation. In Fig.~\ref{fig4}(c), we show the Fermi surface which consists of two large sheets. Below the structural transition temperature, the system has a lower symmetry\cite{Slebarski2013}. This corresponds to a reduction in the size of the first Brillouin zone and the large Fermi surfaces will be gapped out, resulting in a smaller density of states at the Fermi energy. The calculation using the high temperature structure thus gives an upper bound to $\gamma_{\rm{cal}}$. Comparing with the value obtained from the specific heat analysis above, $\gamma=27.87$~mJ$\cdot$K$^{-2}$mol$^{-1}$ obtained from experiment is larger than $\gamma_{\rm{cal}}$ and one could attribute the excess in $\gamma$ to the contribution from the electron-phonon coupling and/or electron-electron interaction. In addition, the Fermi surface consists of areas of low curvatures. This is a scenario which may favor nesting and enhance the electron pairing in the superconducting state.

In summary, we have measured the electrical resistivity, the specific heat and the X-ray diffraction of \LaSn. The low temperature normal state data follow the Fermi liquid behaviour. A second-order phase transition is unambiguously established at $T^*=152$~K through a careful examination of the temperature evolution of our data. The existence of lattice instability is supported by our lattice dynamic calculations where imaginary phonon frequencies are found near ${\bf M}$, which corresponds to $\bm{q}=(0.5, 0.5, 0.0)$, when the high temperature structure with the space group $Pm\bar{3}n$ is assumed. The electronic band structure calculations result in two Fermi surface sheets which consist of low curvature segments. These results point to strong similarities between \LaSn\ and \SrRh /\SrIr, and establish \LaSn\ as another possible system to investigate the interplay between structural instability and superconductivity.

\begin{acknowledgments} This work was supported by Research Grant Council of Hong Kong (ECS/24300214, ECS/24300814), CUHK Direct Grant (No. 4053123, No. 3132747), Grant-in-Aids for Scientific Research (A) (No. 23244068), (B) (No. 16H04131), and (C) (No. 24540336) from Japan Society for the Promotion of Science, and National Science Foundation China (No. 11504310).
\end{acknowledgments}



\end{document}